\documentclass{acm_proc_article-sp}
\usepackage[latin1]{inputenc}
\usepackage{url}
\usepackage{graphicx}
\usepackage[tight]{subfigure}
\usepackage{multirow}

\begin{document}

\title{A Flexible Thread Scheduler for Hierarchical Multiprocessor Machines}

\numberofauthors{1}
\author{\alignauthor Samuel Thibault \\
	\affaddr{LaBRI}\\
	\affaddr{Domaine Universitaire, 351 cours de la libération}\\
	\affaddr{33405 Talence cedex, France}\\
	\email{samuel.thibault@labri.fr}\\
}

\date{19 June 2005}
%\date{%\small March 15th, 2005}

%\conferenceinfo{COSET-2}{'05 Cambridge, Massachusetts USA}
%\CopyrightYear{2005}

\maketitle

\begin{abstract}
With the current trend of multiprocessor machines towards more and more
hierarchical architectures, exploiting the full computational
power requires careful distribution of execution threads and data so as to
limit expensive remote memory accesses. Existing multi-threaded
libraries provide only limited facilities to let applications
express distribution indications, so that programmers end up
with explicitly distributing tasks according to the underlying
architecture, which is difficult and not portable.
In this article, we present: (1) a model for
dynamically expressing the structure of the computation; (2) a
scheduler interpreting this model so as to make judicious
hierarchical distribution decisions; (3) an implementation within the
\textsc{Marcel} user-level thread library. We experimented our proposal
on a scientific application running on a ccNUMA \textsc{Bull}
\textsc{NovaScale} with 16 \textsc{Intel} \textsc{Itanium} II
processors; results show a 30\% gain compared to a classical scheduler,
and are similar to what a handmade scheduler achieves in a
non-portable way.
\end{abstract}
%\keywords{threads, schedule, NUMA, SMP, Multi-Core}

\section{Introduction}

``Disable HyperThreading!'' That is unfortunately the most common
pragmatic answer to performance losses noticed on HyperThreading-capable
processors such as the \textsc{Intel} \textsc{Xeon}. This is of particular
concern since hierarchy depth has increased over
the past few years, making current computer architectures more and
more complex (\textsc{Sun WildFire} \cite{WildFire}, \textsc{Sgi
Origin} \cite{Origin}, \textsc{Bull NovaScale} \cite{NovaScale} for
instance).

Those machines look like Russian dolls: nested technologies
allow them to execute several threads at the same time on the same
core of one processor (SMT: Simultaneous Multi-Threading),
to share cache memory between several cores (multicore chips), and
finally to interconnect several multiprocessor boards (SMP) thanks to crossbar networks. The
resulting machine is a NUMA (Non-Uniform Memory Architecture) computer, on which the memory access delay depends on the
relative positions of processors and memory banks (this is called the
``NUMA factor'').

The recent integration of SMT and multicore technologies
make the structure of NUMA machines even more complex, 
yet operating systems still have not exploited previous NUMA machines
efficiently. Hennessy and Patterson underlined that fact \cite{CompArch}
about systems proposed for \textsc{SGI Origin} and \textsc{Sun
Wildfire}: \emph{``There is a long history of software lagging behind on
massively parallel processors, possibly because the software problems
are much harder.''} The introduction of new hardware technologies
emphasizes the need for software development.  Our goal is to provide a
\emph{portable} solution to enhance the efficiency of high-performance
multi-threaded applications on modern computers.

Obtaining optimal performance on such machines is a significant challenge.
Indeed, without any information on tasks' affinity,
it is difficult to make good decisions about how to group
tasks working on a common data set on NUMA nodes.
Detecting such affinity is
hard, unless the application itself somehow expresses it.

To relieve programmers from the burden of redesigning the whole task scheduling
mechanism for
each target machine, we propose to establish a communication between the
execution environment and the application so as to automatically get
an optimized schedule. The application describes the organization
of its tasks by grouping those that work on the same data
(memory affinity) for instance. The system scheduler can then exploit this
information by adapting the task distribution to the hierarchical levels of
the machine.

Of course, a universal scheduler that would get good results by using
only such a small amount of information remains to be written. In the
meantime, we provide facilities for applications to query the system about the
topology of the underlying architecture and ``drive'' the scheduler.
As a result, the programmer can easily try and
evaluate different gathering strategies.
%He may even discover affinities he would not have suspected.
More than a mere scheduling model, we propose a scheduling
experimentation platform.

In this article, we first present the main existing approaches that
exploit hierarchical machines, then we propose two new
models describing application tasks and hierarchical levels of the
machine, as well as a scheduler that takes advantage of them. Some
implementation details and evaluation results are given before
concluding.

\section{Exploiting hierarchical\\machines}

Nowadays, multiprocessor machines like NUMAs with multi-threaded
multicores are increasingly difficult to exploit. Several approaches
have been considered.

\subsection{Predetermined distribution and\\scheduling}
\label{manual}

For very regular problems, it is possible to determine a task schedule
and a data distribution that are suited to the target machine and its
hierarchical levels. The application just needs to get the system
to apply that schedule and that distribution, and
excellent (if not optimal) performance can be obtained. The
\textsc{PaStiX}\cite{PaStiX} large sparse linear systems solver is a
good example of this approach.  It first launches a simulation of the
computation based on models of BLAS operators and communications on
the target architecture. Then it can compute a static schedule of
block-computations and communications.

So as to enforce these scheduling strategies, many systems (\textsc{Aix},
\textsc{Linux}, \textsc{Solaris}, \textsc{Windows}, ...) allow
process threads to be bound to processor sets, and memory allocations
to be bound to memory nodes. Provided that the machine is dedicated to
the application, the thread scheduling can be fully controlled by
binding exactly one thread to each processor. To perform task switching,
mere explicit context switches may be used: threads are only used as
execution flow holders.

\subsection{Opportunist distribution and scheduling}
\label{selfsched}

Greedy algorithms (called Self-Scheduling (SS) \cite{ss}) are dynamic,
flexible and portable solutions for loop parallelization. Whatever the
target machine, a Self-Scheduling algorithm takes care of both thread
scheduling and data distribution. Operating systems schedulers are based
on these algorithms.

They basically use a single list of ready tasks from which the scheduler just
picks up the next thread to be scheduled. Hence the workload is
automatically distributed between processors.
%TODO: garder ?
For each task, the last
processor on which it was scheduled is recorded, so as to try to
reschedule it on the same processor as much as possible to avoid cache misses.
These techniques are used in the \textsc{Linux}~2.4 and
\textsc{Windows}~2000 \cite{WinSched} operating systems. However, a
unique thread list for the whole machine is a bottleneck, particularly
when the machine has many processors.

To avoid such contention, Guided Self-Scheduling (GSS) \cite{gss} and
Trapezoid Self-Scheduling (TSS) \cite{tss} algorithms make each
processor take a whole part of the total work when they are idle, raising
the risk of imbalances. AFfinity Scheduling (AFS) \cite{afs} and
Locality-based Dynamic Scheduling (LDS) \cite{lds} algorithms use
a per-processor task list. Whenever idle, a processor will steal work
from the least loaded list, for instance. These latter
algorithms are used by current operating systems (\textsc{Linux}~2.6
\cite{lse},
\textsc{FreeBSD}~5.0 \cite{ule},
\textsc{Cellular Irix} \cite{Irix}).
They also add a few rebalance policies: new processes are
charged to the least loaded processor, for instance.
%the most loaded processor
%periodically gives some of its work to the least loaded one, etc.

However, contention appears quickly with an increased number of
processors, particularly on NUMA machines. \textsc{Wang} \emph{et~al}.
propose a Clustered AFfinity Scheduling (CAFS) \cite{cafs} algorithm
which groups $p$ processors in groups of $\sqrt{p}$. Whenever idle,
rather than looking around the whole machine, processors steal work from
the least loaded processor of their group, hence getting better
localization of list accesses.  Moreover, by aligning groups to NUMA
nodes, data distribution is also localized.  Finally, the Hierarchical
AFfinity Scheduling (HAFS) (\textsc{Wang} \emph{et~al}.  \cite{hafs})
algorithm lets any idle group steal work from the most loaded group.
This latter approach is being considered for latest NUMA-aware
developments of operating systems such as \textsc{Linux}~2.6 and
\textsc{FreeBSD}.

\subsection{Negotiated distribution and scheduling}
\label{NegociatedSched}

There are intermediate solutions between predetermined and opportunist
scheduling. Some language extensions such as \textsc{OpenMP}~\cite{OpenMPIntro},
HPF (\emph{High Performance Fortran})
\cite{hpfintro} or UPC (\emph{Unified Parallel C}) \cite{upc} let one
achieve parallel programming by simply annotating the source
code. For instance, a \texttt{for} loop may be annotated to be
automatically parallelized. An HPF matrix may be annotated to be
automatically split into rather independent domains that will be
processed in parallel.

The distribution and scheduling decisions then belong to the
compiler. To do this, it adds code to query the execution environment
(the number of processors for instance) and compiles the program in a
way generic enough to adapt to the different parallel
architectures. In particular, it will have to handle threads for
parallelized loops or distributed computing, and even handle data
exchange between processors (in the case of distributed
matrices of HPF). To date, expressiveness is limited mostly to ``{\tt Fork-Join}''
parallelism, which means, for instance, that
the programmer can not express imbalanced parallelism.

Programmers may also directly write applications that are able to adapt
themselves to the target machine at runtime. Modern operating
systems provide full information about the architecture of the machine
(user-level libraries are available: \texttt{lgroup} \cite{SolarisLgroup}
for \textsc{Solaris} or \texttt{numa}~\cite{lse} for \textsc{Linux}).
The application can then not only get the number of processors,
but also get the NUMA nodes hierarchy, their respective number of
processors and their memory sizes. Those systems also let the
application choose the memory allocation policy (specific memory node,
\emph{first touch} or \emph{round robin}) and bind threads to CPU sets.
Thus, the application controls threads and memory distribution, but it
is then in charge of balancing threads between processors.

\subsection{Discussion}

We chose to classify existing approaches into three categories.
The \emph{predetermined} category gives excellent performance. But it
is portable only if the problem is regular, \emph{i.e.,} its solving time
depends on the data structure and not on the data itself.
The \emph{opportunist} approach scales well, but does not take
task affinities into account, and thus, on average, does not get
excellent performance.
The \emph{negotiated} approach lets the application adapt itself to the
underlying machine, but requires rewriting of some parts of the scheduler 
in order to be flexible.

Our proposal is a mix between negotiated and opportunist approaches. We
will give the programmers means to dynamically describe how their
applications behave, and use this information to guide a generic
opportunist scheduler.

\section{Proposal: an application-guided scheduler}
\label{contribution}

Our proposal is based on a collaboration between the application and its
execution environment.

\subsection{Bubbles modeling the application\\structure}

The application is asked to model the general layout of its threads
in terms of nested sets called {\bf bubbles}\footnote{In a way relatively
similar to some communication libraries such as MPI, which ask the
application to specify \emph{communicators}: groups of machines which
will communicate.}.

\begin{figure*}
\centering
\includegraphics[scale=0.8]{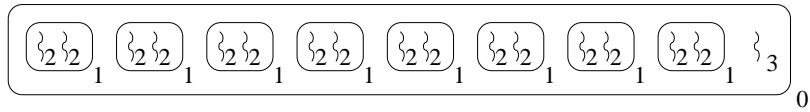}
\caption{\small Bubble example, with priorities: thread pairs that have
a higher priority than the bubbles holding them, and a highly
prioritized thread.}
\label{ex_bubbles}
\end{figure*}

Figure \ref{ex_bubbles} shows such a model: the application groups
threads into pairs, along with a communication thread (priorities will
be discussed later). The concept of bubbles can be understood as a
\emph{coset with respect to a specific affinity relation}, and bubble nesting
expresses refinement of a relation by another one. Indeed, several
affinity relations can be considered, for instance:

\begin{description}
\item[Data sharing] It is a good idea to group threads that work on the
same data so as to benefit from cache effects, or at least to avoid
spreading the data throughout the NUMA nodes thereby incurring the NUMA factor
penalty.
\item[Collective operations]
It can be beneficial to optimize the
scheduling of threads which are to perform collective
operations such as a synchronization barrier, which ensures that
all involved threads have reached the barrier before they can continue
executing.
\item[SMT] Many attempts were made to address thread scheduling on Simultaneous
Multi-Threading (SMT) processors, mostly by detecting
affinities between threads at runtime \cite{HTsched,HtPerfCtrSched}.
Indeed, in some cases,
pairs of threads may be able to efficiently
exploit the SMT technology: they can run in parallel on the logical
processors of the same physical processor without interfering.  If the
programmer knows that some pairs of threads can work in such
\emph{symbiosis}, he can express this relation.
\end{description}
Other relations may be possible to express parallelism, sequentiality,
preemption, etc. Yet, blindly expressing these relations may also be
detrimental: \textsc{Bulpin} and \textsc{Pratt} show performance
loss~\cite{P4HTPerfs} on SMT processors due to frequent cache misses for
instance; \textsc{Antonopoulos} \emph{et~al.} also show performance
loss~\cite{SMPBusLimit} when not taking the SMP bus bandwidth limit into
account.
But the programmer may try and test different refinements of the
relations and thus experimentally reveal how the threads of an
application should be related.

In order to cope with the emerging multiprocessor networks of the 1980's,
\textsc{Ousterhout} \cite{gang} proposed to group data and threads by
affinity into \emph{gangs}. These gangs hold a fixed number of threads which
are to be launched at the same time on the same machine of the
network: this is called \emph{Gang Scheduling}. However, processors may be left idle because a
single machine can only run one gang at a time, even if it is ``small''.
\textsc{Feitelson} \emph{et~al}. \cite{gang_design} propose a
hierarchical control of the processors so as to execute
several gangs on the same machine.
%
%Besides, the \textsc{Solaris} operating system lets applications bind
%threads to LWPs (Light-Weight Processes). The system takes care of
%distributing LWPs on the different processors while the user-level
%thread library can, on each LWP, schedule threads bound to it.
%
Our approach is actually a generalization of
%those.
this approach.

\subsection{Task lists modeling the computing power structure}
\label{runqueues}

According to \textsc{Dandamudi} and \textsc{Cheng} \cite{runqueues}, a
hierarchy of task lists generally brings better performance than simple
per-processor lists. This is why two-level list schedulers have been
developed~\cite{TwoLevelSched,TwoLevelSched2}.
Moreover, it makes task binding
to processor sets easier.
In a manner similar to \textsc{Nikolopoulos} \textit{et al.}'s Nano-Threads list
hierarchy~\cite{NanoSched}, we have taken up and generalized this point of view.

\begin{figure*}[tb]
\centering
\subfigure[A NUMA of HyperThreaded multicores.]
{
\includegraphics[scale=0.6]{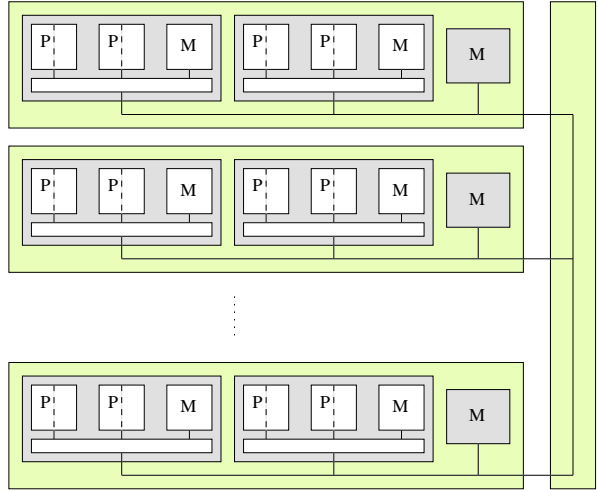}
\label{super-numa-phys}
}
\subfigure[Model with task lists.]{
\includegraphics[scale=0.6]{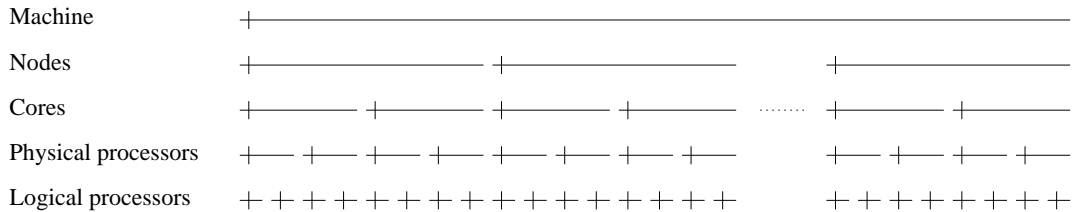}
\label{super-numa-list}
}
\caption{\small A high-depth hierarchical machine and its model.}
\label{super-numa}
\end{figure*}

Indeed, we model hierarchical machines by a hierarchy of task
lists. Each component of each level of the hierarchy of the machine has
one and only one task list. Figure \ref{super-numa} shows a
hierarchical machine and its model. The whole machine, each NUMA
node, each core, each physical SMT processor and each logical SMT
processor has a task list.

For a given task, the list on which it is inserted expresses the
scheduling area: if the task is on a list associated with a physical chip,
it will be allowed to be run by any processor on this chip; if it is
placed on the global list, it will be allowed to be run by any processor
of the machine.

% garder ?
%Any organization can be represented this way. It is even possible to
%model a set of NUMA machines linked up by \arevoir{an
%addressing-capable} network (such as SCI \footnote{An SCI network
%permits to define some memory segments which are transparently shared by
%several machines, the \textsc{Sequent NUMA-Q} machine \cite{NUMA-Q} uses
%that technique.} \cite{sci}): the ``Network'' level just needs to be
%added on top of the machine level.
%%
%Of course, the physical hierarchy needs not be followed.
%For instance, artificial processor clusters that CAFS (seen in
%\ref{selfsched}) proposes can also be modeled.

\subsection{Putting both models together: a bubble scheduler}
\label{bubble_scheduler}

Once the application has created bubbles, threads and bubbles are
just ``tasks'' that the execution environment distributes on the
machine.

\subsubsection{Bubble evolution}
\label{evolution}

\begin{figure*}
\centering
\subfigure[]
{\label{bulle1}\includegraphics[scale=0.75]{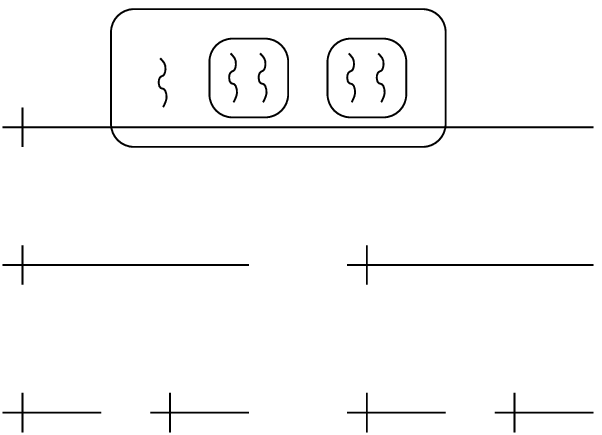}}
\hspace{0.8cm}
\subfigure[]
{\label{bulle2}\includegraphics[scale=0.75]{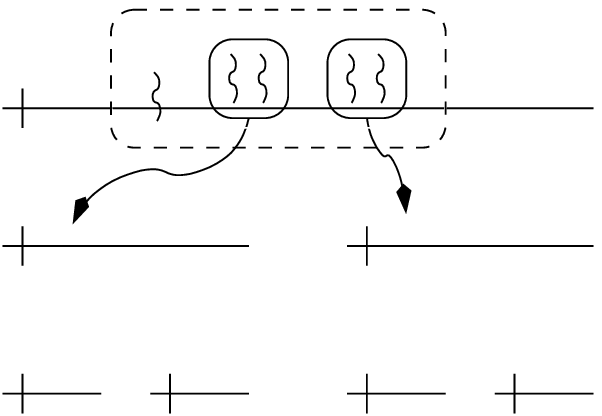}}
\hspace{0.8cm}
\subfigure[]
{\label{bulle3}\includegraphics[scale=0.75]{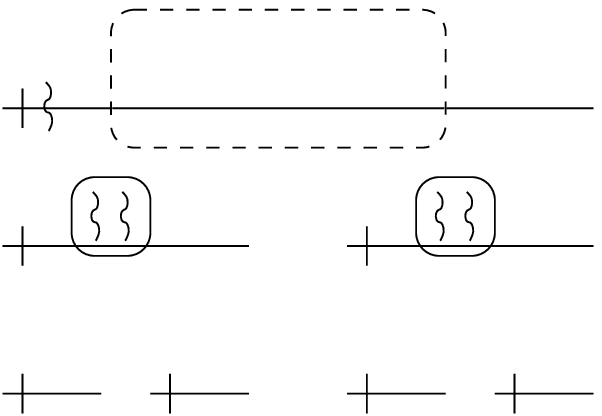}}
\subfigure[]
{\label{bulle4}\includegraphics[scale=0.75]{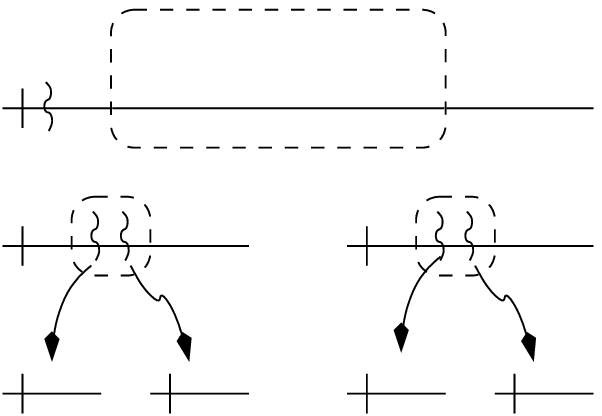}}
\hspace{0.8cm}
\subfigure[]
{\label{bulle5}\includegraphics[scale=0.75]{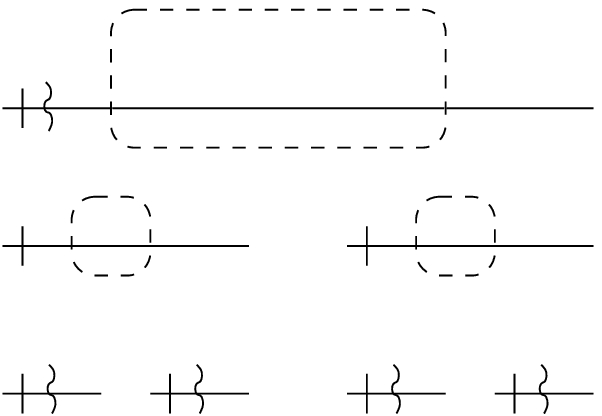}}
\caption{\small Bubble evolution.
{(a) The outermost bubble starts on the general list.
(b) It bursts, releasing a thread  (which can immediately be scheduled on
any processor) and two sub-bubbles which can go down through the hierarchy.
(c) Going down achieved.
(d) Both sub-bubbles burst, releasing two threads each.
(e) Threads are distributed appropriately and can start in parallel.}
}
\label{bubbles_evolution}
\end{figure*}

As Figure~\ref{bubbles_evolution} shows, the goal of a bubble is to hold
tasks and bring them to the level where their scheduling will be
most efficient. For this, the bubble goes down through lists to the
\emph{wanted} hierarchical level.  It then ``bursts'', \emph{i.e.} held
threads and bubbles are released and can be executed (or go deeper).
The list of held tasks is recorded, for a potential later
regeneration (see Section~\ref{re-formation}).
The main issue is how to specify the right bursting level of a bubble.

In the long run, once we get good heuristics for a bubble
scheduler, specifying such a parameter will no longer be necessary. For
now, the goal is to provide an experimental platform for developing
schedulers, and hence allow this parameter to be tuned by the scheduler
developers.
% garder ?
%For this, he basically sets for each bubble its bursting level, as an
%integer between 1 and a chosen $n$. Our scheduler then linearly
%distributes these $n$ virtual levels on the real levels of the machine,
%so that virtual level 1 maps to the ``whole machine'' real level and
%virtual level $n$ maps to the ``logical processor'' real level.
They can favor task affinity with the risk of
making the load balance difficult (by setting deep bursting levels) or on the
contrary favor processor use (by setting high bursting levels).

\subsubsection{Priorities}
\label{prios}
% garder ? Attention aux références un peu partout !

We choose to let the application attach integer priorities to tasks.
When a processor looks for a task to be scheduled, it
searches through the lists that ``cover'' this processor, from the most
\emph{local} one (\emph{i.e.} on low levels) to the most \emph{global}
one,
looking for a task with highest priority. It will then schedule that
task, even if less prioritized tasks remain on more local lists.

Figure \ref{ex_bubbles} shows an example using priorities. In this example,
bubbles holding computing threads are \emph{less} prioritized than the
threads.  Consequently,
a bubble will burst only if every thread of the
previously burst bubbles has terminated, or if there are not enough
of them to occupy all the processors. This results in some \emph{Gang
scheduling} which automatically occupies all the processors.

\subsubsection{Bubble regeneration}
\label{re-formation}

Bubbles are automatically distributed by the scheduler over the
different levels of task lists of the machine, hence distributing
threads on the whole machine while taking affinity into account.
However, it is possible that a whole thread group has far less work
than others and terminates before them, leaving idle the whole part of
the machine that was running it.

To \emph{correct} such imbalance, some bubbles may be regenerated and
moved up. Idle processors would then move some of them down on their
side and have them re-burst there, getting a new distribution
suited to the new workload while keeping affinity intact.

To \emph{prevent} such imbalances, bubbles may periodically be
regenerated\footnote{In a way similar to \textsc{Unix} system thread
preemption.}: each bubble has its own \emph{time slice} after which its
threads are preempted and the bubble regenerated.

%: garder ?
In the case of Figure~\ref{ex_bubbles}, the preemption mechanism is
extended to \emph{Gang Scheduling}: whenever a bubble is regenerated
(because its time slice expired), it is put back at the end of the task
list while another bubble is burst to occupy the resulting idle
processors.

\subsection{Discussion}

Bubbles give programmers the opportunity to express the structure of
their application and to guide the scheduling of their threads in a
\emph{simple}, \emph{portable} and \emph{structured} way. Since the roles of
processors and other hierarchical levels are not predetermined, the
scheduler still has some degrees of freedom and can hence use an opportunist
strategy to distribute tasks over the whole machine. By taking into
account any irregularity in the application, this scheduler
significantly enhances the underlying machine exploitation.
Such preventive rebalancing techniques may still have side effects and
lead to pathological situations (ping-ponging between tasks, useless bubble
migration just before termination, etc.).

\section{Implementation details}

\textsc{Marcel}~\cite{pm2,Dan} is a two-level thread library: in a way
similar to manual scheduling (see section \ref{manual}), it binds one
kernel-level thread on each processor and then performs fast user-level
context switches between user-level threads, hence getting complete
control on threads scheduling\footnote{We suppose that no other
application is running, and neglect system daemons wake-ups.}
in userland without any further help from the kernel. Our
proposal was implemented within \textsc{Marcel}'s user threads scheduler.

Figure \ref{bubble-code} shows an example of using the interface to
build and launch a bubble containing two \textsc{Marcel} threads.

\begin{figure}
\centering
{\small\begin{verbatim}
   marcel_t thread1, thread2;
   marcel_bubble_t bubble;

   marcel_bubble_init(&bubble);
   marcel_create_dontsched(&thread1, NULL, fun1, para1);
   marcel_create_dontsched(&thread2, NULL, fun2, para2);
   marcel_bubble_inserttask(&bubble, thread1);
   marcel_wake_up_bubble(&bubble);
   marcel_bubble_inserttask(&bubble, thread2);
\end{verbatim}}
\caption{\small Bubble creation example: threads are created without
being started, then they are inserted in the same bubble.}
\label{bubble-code}
\end{figure}

The \textsc{Marcel} scheduler already had per-processor thread lists,
so that integrating
bubbles within the library did not need a thorough rewriting of the data
structures. The scheduler code was modified to implement
list hierarchy, bubble evolution and to take priorities (described in
Section~\ref{prios}) into account.

So as to avoid contention, there is no global scheduling: processors
just call the scheduler code themselves whenever they preempt (or
terminate) a thread. The scheduler finds some thread that is ready to be
executed by the processor. We added bubble management there: while looking for
threads to execute, the scheduler code now also tries to ``pull down''
bubbles from high list levels and make them burst on a more local level.
Getting an efficient implementation is complex, as explained below.

Given a processor, two passes are done to look for the task (thread or
bubble) with maximum priority among all the tasks of the lists
``covering'' that processor. The first pass quickly finds the list
containing the task with the highest priority, without the need of a lock. That
list and the list holding the currently running task are
locked\footnote{By convention, locking lists is done by locking high-level
lists first, and for a given level, according to the level elements
identifiers.}.
A second pass is then used to check that the selected list
still has a task of this priority, in case some other processor took it
in the meantime. If the selected task is a thread, it is scheduled;
otherwise it is a bubble that the processor deals with appropriately
(going down / bursting). The implementation time-complexity is
linear with respect to the number of hierarchical levels of the machine.

Regenerating a bubble is also a difficult operation. Replacing
threads in a given bubble requires removing all of them from the task
lists, except threads being executed. Those threads go back in the
bubble by themselves when the processors executing them call the
scheduler. Eventually, the last thread closes the bubble and moves it
up to the list where it was initially released by the bubble holding it.

\section{Performance evaluation}

Our algorithm has some cost, but increases performance thanks to the
resulting localization.

\subsection{Bubble scheduler cost}

We measured the performance impact of our implementation on the
\textsc{Marcel} library running on a 2.66\,GHz \textsc{Pentium IV Xeon}.
Searching through lists has a reasonable cost, and our scheduler
execution times are good compared to the \textsc{Linux} thread libraries
\textsc{LinuxThread} (2.4 kernel) and NPTL (2.6 kernel), see Table~\ref{microbenchmark}.

\begin{table}
\small
\centering
%\begin{tabular}{|l|c|c|c|}
%\cline{3-4}
%\multicolumn{2}{l|}{ }& Time (ns) & $\sim$ cycles \\\hline
%\multirow{2}{*}{Context switch only}
%& \texttt{setjmp} & 7.1 & 19 \\\cline{2-4}
%& \texttt{longjmp} & 5.6 & 15 \\\hline
%\multirow{2}{*}{\textsc{Marcel}}
%& \texttt{yield} & 250 & 665 \\\cline{2-4}
%& \texttt{yield} + switch & 398 & 1060 \\\hline
%\multirow{2}{*}{\textsc{NPTL}}
%& \texttt{yield} & 672 & 1790 \\\cline{2-4}
%& \texttt{yield} + switch & 2150 & 5720 \\\hline
%\multirow{2}{*}{\textsc{LinuxThread}}
%& \texttt{yield} & 672 & 1790 \\\cline{2-4}
%& \texttt{yield} + switch & 1730 & 4600 \\\hline
%\end{tabular}
\begin{tabular}{|l|c|c|c|c|c|c|}
\cline{2-7}
\multicolumn{1}{l|}{}&\multicolumn{3}{|c|}{Yield}&\multicolumn{3}{|c|}{Switch}\\
\cline{2-4}\cline{5-7}
\multicolumn{1}{l|}{}&ns&cycles&\%&ns&cycles&\%\\\hline
\textsc{Marcel} (original)&186&495&69&84&223&31\\ \hline
\textsc{Marcel} bubbles   &250&665&63&148&395&37\\ \hline
NPTL (Linux 2.6)          &672&1790&31&1488&3930&69\\ \hline
\end{tabular}
\caption{\small
Cost of the modified \textsc{Marcel} scheduler
for searching lists, compared to other schedulers.
Yield: list search only, Switch:
synchronization and context switch.
}
\label{microbenchmark}
\end{table}

Creation and destruction of a bubble holding a thread does not
cost much more than creation and destruction of a simple thread: the cost
increases from 3.3$µ$s to 3.7$µ$s.

\begin{figure}[tb]
\centering
\subfigure[2 HyperThreaded \textsc{Pentium IV Xeon}]
{\includegraphics[scale=0.62]{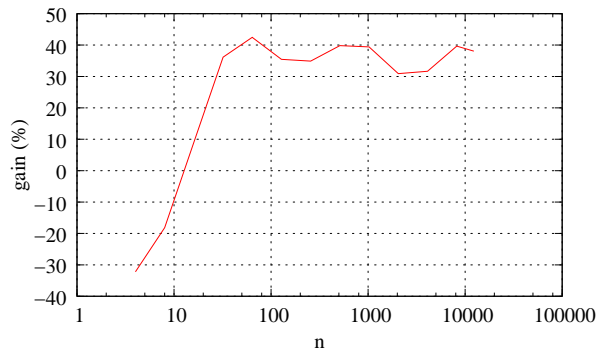}}
\subfigure[$4\times4$ \textsc{Itanium II}]
{\includegraphics[scale=0.62]{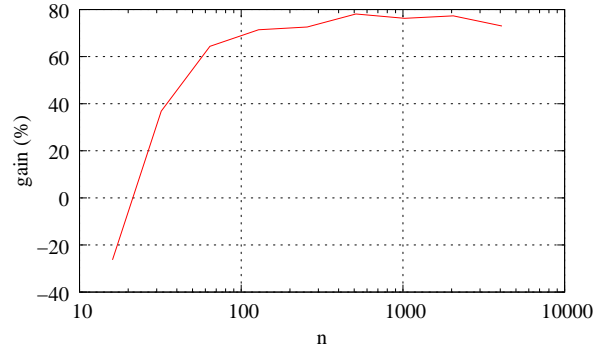}}
\caption{\small Performance gain brought by adding bubbles to the
\emph{fibonacci}
test-case.}
\label{sumtime}
\end{figure}

Test-case examples of recursive creation of threads, such as
divide-and-conquer \emph{Fibonacci} show that the cost of
systematically adding bubbles that express the natural recursion of threads creations is
quickly balanced by the localization that they bring:
Figure~\ref{sumtime} shows that performance is affected when only a few
threads are created, while on a HyperThreaded Bi-\textsc{Pentium IV
Xeon}, the performance gain stabilizes at around 30 to 40\% with 16 threads;
on a NUMA $4\times4$ \textsc{Itanium II}, the gain is 40\% with 32 threads
and gets up to 80\% with 512 threads.

\subsection{A real application}

%During his PhD at CEA DAM, 
Marc \textsc{Pérache}~\cite{MarcRenPar05} used our
scheduler in a comparison of the efficiency of
various scheduling strategies for \emph{heat conduction and advection} simulations.
Results may be seen in Table~\ref{conduction}. The target machine is a
ccNUMA \textsc{Bull NovaScale} with 16 Itanium II processors and 64\,GB of
memory, distributed among 4 NUMA nodes.  For a given processor,
accessing the memory of its own node is about 3 times faster than
accessing the memory of another node. The applications perform cycles of
fully parallel computing followed by global hierarchical communication
barrier.

In the \emph{simple} version, the mesh is split into as many stripes as
the number of processors, and an opportunist schedule is used.  The
\emph{bound} version binds them to processors in a non-portable way.
This gets far better performance: each thread remains on the same node,
along with its data.  Our proposal lets the application query
\textsc{Marcel} about the number of NUMA nodes and processors and then
automatically build bubbles according to the hierarchy of the machine
(hence 4~bubbles of 4~threads in this example).
It gets performance very
similar to those of the \emph{bound} version.

\begin{table}
\small
\centering
\begin{tabular}{|l|c|c|c|c|}
\cline{2-5}
\multicolumn{1}{l|}{ }&
\multicolumn{2}{|c|}{Conduction}&
\multicolumn{2}{|c|}{Advection}\\
\cline{2-5}
\multicolumn{1}{l|}{ }& Time (s) & Speedup & Time (s) & Speedup\\ \hline
Sequential              & 250.2 &        & 16.13 &       \\ \hline
Simple  & 23.65 & 10.58  & 1.77  & 9.11  \\ \hline
Bound   & 15.82 & 15.82  & 1.30  & 12.40 \\ \hline
Bubbles & 15.84 & 15.80  & 1.30  & 12.40 \\ \hline
\end{tabular}
\caption{\small \textsc{Conduction} performance depending on the approach.}
\label{conduction}
\end{table}

As can be seen,
the use of bubbles attained performance close to that which may be
achieved with a ``handmade'' thread distribution, but in a portable way.

These applications are a simple example in which
the workload is balanced between stripes. The use of bubbles simply allowed it to
automatically fit the architecture of the machine. However, in the
future these
applications will be modified to benefit from Adaptive Mesh Refinement
(AMR) which increases computing precision on interesting areas. This
will entail large workload imbalances in the mesh both \emph{at runtime} and
\emph{according to the computation results}. It will hence be
interesting to compare both development time and execution time of handmade-,
opportunist-, and bubble-scheduled versions.

\section{Conclusion}

Multiprocessor machines are getting increasingly hierarchical. This makes
task scheduling extremely complex. Moreover, the challenge is to
get a scheduler that will perform ``good'' task scheduling on any
multiprocessor machine with an arbitrary hierarchy, only
guided by \emph{portable} scheduling hints.

In this paper, we presented a new mechanism making significant progress in that
direction: the bubble model lets applications express affinity relations
of varying degrees between tasks in a portable way. The scheduler can
then use these hints to distribute threads.

Ideally, the scheduler would need no other information to perform
this. But practically speaking, writing such a scheduler is
difficult and will need many experiments to be tuned. In the meantime,
the programmer can use stricter guiding hints (indicating bubble
bursting
levels, for instance) so as to experiment with several strategies.

Performance observations on several test-cases are promising, far better
than what opportunist schedulers can achieve, and close to what
predetermined schedulers get. These observations were obtained on
several architectures (\textsc{Intel} PC SMP, \textsc{Itanium} II NUMA).

This work opens numerous future prospects. In the
short term, our proposal will be included within test-cases of real applications of
CEA that run on highly hierarchical machines, hence stressing the bubble
mechanism power. It will then be useful to develop analysis tools based
on tracing the scheduler at runtime, so as to check and refine scheduling
strategies. It will also be useful to let the programmer set other
attributes than just priorities, and thus influence the scheduler:
``strength'' of the bubble (which expresses the amount of affinity that
the bubble represents), preemptibility, some notion of amount of work, ...

In the longer term, the goal is to provide a means of expression
powerful and portable enough for the application to obtain an automatic
schedule that gets close to the ``optimal'' whatever the underlying
architecture. It could also be useful to provide more powerful memory
allocation functions, specifying which scope of tasks (a bubble for
instance) will use the allocated area.

\bibliographystyle{abbrv}
\bibliography{coset}
%TODO à mettre !
%\balancecolumns
\end{document}